# High temperature induced B1 to B2 phase transformation in Cd doped ZnO nanocomposite thin films

**Arkaprava Das[1,*], C.P. Saini[1]**

[1]Inter University Accelerator Centre, Aruna Asaf Ali Marg, New Delhi-110067, India

## Abstract

In this paper, we report the high temperature mediated structural phase transformation (PT) from B1 (NaCl) to B2 (CsCl) phase in solgel derived $Cd_{0.2}Zn_{0.8}O$ nanocomposite thin film for the temperature (T) range of ~ 700 to ~ 900℃. X-ray diffraction (XRD) pattern provides a direct and unambiguous indication of phase evolution with increase in annealing T from 700 to 900℃ along with co-existence of wurtzite phase throughout the whole transformation regime. In fact, significant reduction in lattice parameters in case of transformed B2 phase, extracted by Rietveld refinement provides a direct indication of chemical pressure inducement inside the lattice owing to anticipated mismatch in thermal expansion coefficient which leads to facilitate such PT. Raman spectrum for 900℃ annealed ones, moreover, shows a distinguishable phonon mode at 866 $cm^{-1}$, suggesting the formation of B2 phase at higher T, consistent with XRD results. Scanning electron microscopy (SEM) reveals the agglomeration of nanocrystallites followed by distinct void formation, indicating the evolution of chemical pressure inside the lattice. Soft X-ray absorption (SXAS) measurement reflects a deviation of the inflection point of the absorption edge which implies Fermi level pinning with transforming phase.

---

**Author for correspondence:* arkaprava@iuac.res.in (Dr. Arkaprava Das, Research Scholar at Inter University Accelerator Centre)

## Introduction:

Since last few decades transition metal oxides have attracted a significant amount of attention regarding its interesting aspects with phase stability and compressibility properties at different ambient condition. [1,2] Among various transition metal oxides, CdO and ZnO is attractive targets for wide range of applications including photovoltaics, electrode for storage batteries and so on [3]. At room temperature and pressure CdO, ZnO exist in rocksalt and wurtzite phase respectively [3]. Wurtzite crystal structured ZnO, with a direct band gap of 3.3 eV [4,5] is now being extensively used as potential materials for light emitting diodes [6,7] mechanical energy harvesting device [8]. The high photoluminescence and piezoelectricity are the prime reason behind its potential candidature. CdO and ZnO both are well known transparent conducting oxide (TCO) and massively utilised for opto-electronic applications. Both of the oxides exhibit transparency around the visible region of electromagnetic spectrum [9,10]. In these TCO's doping beyond thermodynamic solubility limit, mixed phase evolution in nanocomposite system is not unusual. With increasing doping percentage in $Cd_xZn_{1-x}O$ pseudo binary alloy system, not only band gap gets tuned to cover the maximum region of the visible spectrum [11] but also changes the electron concentration and mobility [3]. Aforementioned properties make this nanocomposite system ideal and deserving candidate for transparent conducting, light emitting and other opto-electronic applications [5,12].

Structural phase transforming materials are reported as potential active materials in Random Access Memory (RAM) devices, optical phase change storage technology (rewritable compact discs-DVDs, CDs etc) [13,14] and in gas sensing devices [15,16]. Most ionic IIB-VIA compound, CdO, exists in rocksalt (NaCl) structure and exhibits a PT to a higher coordination under high hydrostatic pressure. Theoretical studies showed that 3*d* and 4*d* transition metal monoxide i.e. ZnO and CdO, undergo B1 to B2 PT at 256 and 89 GPa pressure [17,18] and CdO can exist in six different high pressure phases [17] though all phases are not experimentally reported yet. High pressure induced PT from B1 to B2 type structure for CdO is experimentally observed at pressure of 90.6 GPa (hydrostatic pressure) by using diamond anvils cell (DAC) technique [19]. Recently, it has been reported by Chen *et al.* [3] that $Cd_xZn_{1-x}O$ thin film, prepared by radio frequency (RF) sputtering method, suffers a wurtzite (B4) to rocksalt (B1) PT with DAC technique. Most of the literature regarding structural PT, report experiments carried out using DAC technique.

For the first time we are reporting the NaCl to CsCl i.e. B1 to B2 phase transformation for Cd doped ZnO solgel derived thin film which has been induced with due to chemical pressure generated inside the lattice. In particular, by gradual increment of Cd doping in ZnO for investigating its different physico-chemical properties for this pseudo binary alloy system, at a certain particular compositional juncture i.e. $Cd_{0.4}Zn_{0.6}O$, an interesting structural PT is observed. However, systematic investigation is essential to make out this chemical pressure mediated transformation kinetics. In this regards XRD and Raman spectroscopy technique are the trustworthy tools for identifying the phase of the concerned materials. During experimental investigation, both of the techniques reflect the existence of transformed B2 phase and are complementary to each other. In fact, agglomeration of nanocrystallites with T in conjunction with formation of relative larger void as confirmed by SEM is the origin of chemical pressure inside the lattice which in turn leads to such B2 phase transformation. Moreover, shifting of inflection point of absorption edge provides a clear evidence for Fermi level pinning with transformed phase which has been verified numerically with fitting, by utilising Athena Software Package. As we observe a gradual evolution of B2 phase with increasing annealing temperature, therefore there might be possibility of metastable phases in between the PT. For identifying those phases, a thorough investigation is essential regarding the transformation kinetics. Overall, the plausible correlation in the present manuscript among PT, microscopic and electronic properties would be very much crucial and captivating for rigorous apprehension of the fundamental interaction for its paramount role in optoelectronic applications.

**Experimental:**

40% Cd doped ZnO thin films are prepared with solgel chemical synthesis method using spin coater. As a origin of cadmium and zinc, cadmium acetate dihydrate [$Cd(CH_3COO)_2.2H_2O$] and zinc acetate dihydrate [$Zn(CH_3COO)_2.2H_2O$] were taken. Details of the synthesis and all characterization facilities are explained in the following. For preparing the precursor solution zinc acetate dihydrate was mixed with 2-methoxyethanol ($CH_3$-O-$CH_2$-$CH_2$-$CH_2$-OH) with a sol stabilizer i.e. Diethanolamine (DEA) and kept on stirring with the help of a magnetic stirrer for several hours. Subsequently cadmium acetate dihydrate with proper stoichiometric ratio with desired molarity was mixed with solution and we kept on stirring it for few hours. The solution got transparent after few hours and we kept the transparent solution for gelation upto 72 hours. After the solution preparation was accomplished then we coated it on Si wafer with the help of a spin coater at 2800 rpm (revolution per minute). We coated the thin films 12 times and after each subsequent coating we kept the Si wafer substrate upon hot plate at 200 ℃. Successively 12 times coating was performed to achieve homogenous and uniformly distributed nanoparticles upon Si substrate with thickness at our hearts content. After 12 times coating, we have annealed the thin films in a tubular furnace for 700, 750, 800, 850, 900 ℃ in oxygen environment. The nomenclature of the thin films for different characterizations are provided as CZ700, CZ750, CZ800, CZ850, CZ900 to make it distinguishable according to the different annealing temperature. Subsequent characterizations were performed in those annealed thin films.

X-ray diffraction measurements were performed with Bruker High resolution X-Ray diffractometer system using Cu $K_{\alpha}$ beam. Rutherford Backscattering Spectrometry was done with 2 MeV $He^{+}$ impinging beam with detector kept at $165^{0}$ angle. Scanning electron microscopy measurements have been performed with MIRA/TESCAN at Scientium Analyze Solution, Jaipur. Raman Spectroscopic measurements have been accomplished with Renishaw InVia Raman microscope using Ar ion laser (514.5 nm). Soft X-ray absorption measurements have been performed at BL-01 in Raja Ramanna Centre for Advanced Technology (RRCAT), Indore, Madhya Pradesh.

**Compositional and Structural studies:**

Figure 1(a) shows the Rutherford backscattering spectra for CZ900 and CZ700 thin films. Figure 1(b) and 1(c) show the rump simulation for both of the thin films from which the simulated values of thicknesses are found to be 490$\pm$10 nm and 430$\pm$10 nm for CZ700 and CZ900 respectively. Six different layers were taken inside the Si layer and Cd, Zn, O are diffused with a consistent decreasing linear diffusion gradient. However the amount of diffusion especially for Cd is significantly greater in CZ900 thin film compared to CZ700 thin film. Due to higher annealing temperature diffusivity of Cd increases which are clear from the simulation results and also might be the possible reason for lowered backscattering yield in CZ900 thin film spectra. From the spectra we can observe that Cd yield is much lower than the yield of Zn. The backscattering cross-section depends on the mass of the target atom and the same is higher for Zn compared to Cd [20].

In figure 2(a), X-ray diffractograms are shown for step annealed thin films which reflects that phase is transforming from B1 (NaCl) to B2 (CsCl) phase. Here doping percentage is quite high, beyond solubility limit. So, it is quite obvious that distinguishable peaks for CdO rocksalt phase should be there. For CZ700 thin film B1 phase peaks i.e. for CdO rocksalt phase are situated at 33.0 degree (111) and at 38.3 degree (200). As the structural PT is taking place in the nanocomposite system not for pure system. Therefore, the peaks for wurtzite phase due to ZnO are always present throughout the whole PT regime. The peaks for wurtzite phase are situated at 31.6 degree (100), 34.3 (002), 36.1 degree (101), 47.4 degree (102). So CZ700 thin film is consisted of mixed phase i.e. wurtzite+NaCl. Such kind of B1 to B2 PT is reported in the literature both experimentally [19] as well as theoretically [17] with hydrostatic pressure. When we increase the annealing temperature with a step of 50℃, we found a gradual phase evolution around 32 degree. With increasing annealing temperature, that phase becomes quite prominent and for CZ900 thin film, we observe a sharp peak for B2 phase i.e. CsCl phase at 32.2 degree (100) which provides a succinct signature of structural PT to CdO B2 phase. Apart from that peak, the other peaks situated at 37.8 (110), 42.2 degree (111), 45.5 degree (200) also endorse the B2 PT. The ZnO wurtzite phase peaks exist for CZ900 X-ray diffraction pattern. Therefore, CZ900 thin film is subsisted of wurtzite+CsCl phase. From this observation, it is quite clear that PT is taking place in CdO phase inside the host lattice. With increasing annealing temperature the peak position of (100) B2 phase keeps on shifting towards higher $2\theta$ value which gives an indication that compressional stress is generating with increasing annealing temperature. The solgel derived thin films belong to polycrystalline type. Therefore, multiple diffraction peaks from different planes are anticipated and also the formation of secondary phase due to infusing of two oxides is expected [21].

The lattice parameters and space group of the thin films have been further determined for pure ZnO (PZnO), CZ700 and CZ900 with Rietveld method (profile matching with constant scale factor) using FULL PROF SUITE software as shown in figure 2(b), 2(c) and 2(d) for PZnO, CZ700, CZ900 thin films. The refinement of PZ was carried out using space group P63m and the lattice parameters estimated from this sample was used as reference for next two samples. An initial look at figure 2a gives a clear idea that sample CZ700 and CZ900 has character of B1 phase and transformed B2 phase along with wurtzite phase respectively. So, in accordance with that, a two phase model of a cubic (B1) phase (Fm3m) and a Hexagonal phase (P63mc) is used as the refinement strategy for CZ700 and cubic (B2) phase (Pm-3m) and a Hexagonal phase (P63mc) for CZ900 respectively. A Thompson cox hastings function with axial divergence symmetry is used to model peak profiles. The background is fitted using a sixth order polynomial. The refined data (shown in Figure 2(c)) matches quite well with experimental data for CZ700 with GOF ($\chi^2$) =1.88. The fit between the observed and calculated profiles is satisfactory and indicates the correctness of model.

But for the CZ900, there is slight mismatch of the experimental and calculated data which might be due to some impurity peaks present in the samples positioned at 38.8 degree and 41.1 degree. The refined lattice parameters, GOF ($\chi^2$) and cell volume calculated from Rietveld refinement are given in Table 1. We observe that there is a considerable change in the lattice parameters of B1 and B2 phase for CZ700 and CZ900 respectively. i.e. the lattice parameters for CZ900 are nearly half of CZ700. This clearly indicates that with rise in annealing temperature a considerable amount of shrinkage in the volume of the lattice is taking place and thus leading to B1 to B2 phase transition in the system under study.

It is quite clear that from the present scenario, that this PT has not been taken place due to externally executed hydrostatic pressure like the reported literature using diamond anvils cell (DAC) technique [19]. It is taking place due to the generation of chemical pressure with increasing annealing temperature which agrees well with our refinement results. Such kind of lattice chemical pressure induced PT has been reported by Liu *et al.* for $Na_{0.5}K_{0.5}NbO_3$ based ceramic system [22]. In that experimental work they have accomplished ionic substitution by prolonging ball milling time which further generates lattice chemical pressure and gradually transforms the phase. Same kind of PT has been reported by Ahart *et al.* in $PbTiO_3$ system [23]. However, for nano-composite thin film system such kind of chemical pressure induced PT is not reported in the literature to the best of our knowledge. To find the actual value of thermal expansion coefficient for both CdO and ZnO is quite difficult at our transformation temperature regime. However, we know thermal expansion coefficient is an explicit function of temperature. For both CdO and ZnO thermal expansion coefficients increase with increasing temperature. [24–26] As we observe that lattice parameter for CZ900 is becoming almost half for transformed B2 phase compared to B1 phase, therefore increment in the thermal expansion coefficient would be more for ZnO compared CdO. Due to this mismatch in thermal expansion coefficient, ZnO thermally expands and puts pressure upon CdO lattice which transforms the phase for CdO.

If the doping percentage would have been lower, straightforward ionic substitution would have taken place which might result lattice expansion. But here the doping percentage is quite high and phase segregation is clearly observed. Therefore, straightforward ionic substitution would be hindered due to secondary phase evolution. Subsequently, the c axis bond length reduces which provides a signature of generating lattice chemical pressure inside the lattice. CdO (3806 KJ/mol) is having lesser lattice energy compared to ZnO (4142 KJ/mol) [27] and also CdO having melting temperature around 900-1000 ℃. Therefore at 900 ℃ annealing temperature re-evaporation of Cd might take place which might endorse in reducing c axis length.

**Micro-Raman Studies:**

In figure 3 Raman spectra for CZ700, CZ800, CZ900 thin films have been presented. In the present investigation Raman spectroscopic measurements have been performed in backscattering geometry without any polarization consideration. From the spectral observation one can enunciate that CZ700 and CZ800 thin films are CdO dominated whereas CZ900 thin film is ZnO phonon mode dominant. CdO, rocksalt i.e. NaCl structure is having space group Fm$\bar{3}$m. The space group of hexagonal wurtzite phase for ZnO belongs to $C_{6V}^{4}$ where all atoms engross $C_{3V}$ sites in unit cell. There is a well established equation provided by the group theoretician which describes the optical phonon symmetry at the centre of the Brillouin zone. The equation is as follows:

$\Gamma_{opt} = 1A_1 + 2B_1 + 1E_1 + 2E_2$

It is an entrenched phenomena that $A_1$ and $E_1$, are Raman as well as infra-red (IR) active branches. These branches are consisted of polar symmetries which bifurcate itself into doubly degenerate LO and TO components with distinguishable frequencies. $E_2$ is the non-polar branch which is only Raman active not IR active with two $E_2$ (H) and $E_2$ (L) sub-branches. B1 branches are also Raman active and IR inactive like $E_2$ branches [28].

For CZ700 thin film, we observe a broad hump at 276 cm$^{-1}$ and at 949 cm$^{-1}$ which are nothing but the transverse optical (TO) and longitudinal optical (LO) phonon modes for CdO. CdO lattice has a well reported LO and TO phonon modes at 952 cm$^{-1}$ and at 262 cm$^{-1}$ respectively [29,30]. Therefore, our experimental observation completely matches with theoretically reported results. Traditionally in CdO, second order Raman scattering is expected. However, in present case the spectroscopic measurements have been performed with 514.5 nm wavelength Ar laser and this energy is close to the band gap value of CdO. Therefore probability of first order scattering cannot be rules out completely. As CZ700 and CZ800 thin films are CdO phonon mode dominated. Therefore, the vibration of Cd and O sub-lattice are going to influence the $E_2$ (high) and $E_2$ (low) modes which are nonpolar and sensitive to strain. These modes are not influenced by the crystal field. [31] With increasing annealing temperature lattice tends to get more relaxed as they get sufficient duration for relocating the atomic position to minimize the overall entropy of the system. And this relocation, usually results in a phonon softening of strain sensitive phonon modes. Due to such phonon softening red shift of the phonon modes takes place [32]. But interestingly, in our case opposite phenomena is taking place i.e. for CZ800 thin film blue shift is taking place for both 276 cm$^{-1}$ and at 949 cm$^{-1}$ TO and LO mode respectively. Inducement of pressure inside the lattice might be the possible reason behind this anomalous behaviour. As we are increasing annealing temperature the system is tending towards a structural PT. Now for CZ900 thin films we observe a completely changed ZnO mode dominated Raman spectra. Various Raman modes are reported in the literature for ZnO which we observe clearly in our CZ900 thin film spectra. In the literature this has been reported that $A_1$(TO) mode is at 382 $cm^{-1}$, $A_1$ (LO) mode is at 576 $cm^{-1}$, $E_2$ (high) mode is at 438 $cm^{-1}$ [33–35,5]. All these modes are visible in our experimental observation. Particularly the mode, situated at 866 $cm^{-1}$ is corresponding to the transformed B2 phase which has not been reported in the literature so far to the best of our knowledge. Therefore both ZnO wurtzite and cubic CsCl phase are observed to coexist which is compatible with early discussions. The existence of this peak justifies our claim very much that structural PT has been taken place and commensurate with our X-ray diffraction outcome.

**Microscopy study:**

In figure 4(a), (a'), (b), (b'), (c), (c') the SEM images are shown for CZ700, CZ800, CZ900 thin films in 1 μm and 100 nm scale respectively. From the images, it quite clear that nanocrystallites growth has been taken place with increasing annealing temperature. As we know that nanocrystallites growth occurs via two basic processes i.e. cluster migration and Ostwald ripening process. The first one takes place whenever the annealing temperature is below 500°C, the small nanocrystallites start to achieve enough kinetic energy to start coalescence process. When the annealing temperature is beyond 500°C which is our present case Ostwald ripening process comes into the picture. In that particular process larger nanocrystallites construction is ensued with the endorsement of smaller crystallites. Basically, this process is having dependence upon thermal energy available in the system [36]. It follows the higher solubility based Gibbs Thompson rule. According to that rule with increasing annealing temperature, there is a variation of chemical potential at the interface of the nanocrystallites and due to that variation, a positive interfacial energy appears into the picture. That energy endorses the generation of new nanocrystallites with high curvature.

SEM images for the all the thin films reflects an uniform and dense morphology without any ambiguous crack or surprising features. From the observed surface morphology, it is well manifested that there is a homogenous distribution of inter-connected grains with voids throughout the film surface. The voids are becoming quite prominent with increasing annealing temperature as the agglomeration of the nanocrystallites takes place which is quite evident for CZ900 thin film. Traditionally, high temperature annealed and defect free nanocrystallites exhibits more reactivity and subsequently, it would increase the nucleation probability. This enhanced nucleation tendency would help the nanocrystallites to agglomerate [37] and that might be the prime reason behind distinct void formation for CZ900 thin film. However, For CZ900 thin film, at 900℃ annealing temperature which is near about the melting temperature of CdO, there is a probability that oxide gets to start melting and subsequently condensate at energetically assisted sites throughout the whole substrate to minimize the overall surface energy of the system. This would help in generating larger voids in CZ900 thin film.

In previous structural study section, we observe that with PT the volume fraction for CdO phase has been reduced due to generation of pressure. This generation of pressure inside the lattice might be another possible reason for agglomeration of nanocrystallites and prominent void formation.

**Soft X-ray Absorption Spectroscopy (SXAS):**

Figure 5 shows the oxygen *K* edge measurements in total electron yield (TEY) mode for CZ700 and CZ900 thin films. Utilizing Athena XAS data processing software package (code FEFF 6.0) data has been normalized and peak fitting has been performed to dig out the information regarding different peaks which reflects different orbital transitions in the nanocomposite system. From those two spectra, one can observe that there is a significant difference in the spectral features not only at absorption edge bur also beyond 540 eV i.e. beyond absorption edge. Here PT is taking place in CdO phase; therefore we can anticipate that the prime changes in the spectra are coming due to the changed coordination around Cd atom. CdO and ZnO both are n type degenerate semiconductor [38]. Their Fermi level resides beyond conduction band minima (CBM). However, for this nanocomposite system it is quite difficult to depict the actual position of Fermi level but still we can anticipate that it will lie beyond CBM. In oxygen *K* edge for CdO the prime features generates due to the transition from O 1*s* states to empty *p* states which is situated above Fermi level. Generally for transition metal oxides like CdO or ZnO, *s* to *p* orbital transition is dipole allowed which provides a succinct signature that the bonding between anion and cation in those oxide materials are not ionic completely. But they are consisted of mixed character with iconicity and covalence [39]. For completer ionic compound those transitions would have not been permissible as *p* states would have been filled in that case. For our nano-composite system same thing can be anticipated.

During fitting of oxygen *K* edge spectrum for CZ700 thin film, one Arctangent, one Pseudo voigt, three Gaussian and one Lorentzian function were taken. For fitting of the absorption edge, the software took the position of centre of arctangent function i.e. the inflection point of the absorption edge is found to be at 535.72 eV. The other functions are taken at 538.34 eV (Pseudo voigt), 538.34 eV (Gaussian 1), 541.35 eV (Lorentzian 1) and 544.33 eV (Gaussian 2) 547.32 eV (Gaussian 3). For CZ900 thin film the centre of arctangent function is taken 535.62 eV by the Athena software. Apart from that one Gaussian, two Lorentzian and one Pseudo voigt function is taken for fitting which are situated at 539.00 eV (Gaussian), 541.18 eV (Lorentzian 1), 544.60 eV (Pseudo voigt) and 547.02 eV (Lorentzian 2) respectively. Fitted spectra with Athena software have been shown in figure 6. Now, there is a shift of 100 meV in CZ900 thin film with respect to CZ700 thin film in the centre of arctangent function. In core level spectroscopy electrons from core level are jumped to the next unoccupied level. For degenerated semiconductor like CdO, ZnO unoccupancies reside just above Fermi level. Therefore, if the absorption edge is shifting the Fermi level would also get tuned and also the CBM. It is quite obvious that structural PT in any system would change all the ground state properties of the system. This shifting, observed over here from the fitting also provides an indirect evidence that PT is taking place as certainly the position of Fermi level will be different for B1 compared to B2 phase.

For CdO rocksalt structure the CBM is comprised of hybridized orbitals between Cd 5*s* states and O 2*p* orbital states [39]. Therefore, if B1 to B2 structural PT is taking place in CdO phase, the nearest coordination would certainly change which subsequently modify the CBM hybridized states and that change is reflected with modified spectral features in O *K* edge spectra in the A region. The *d* electrons of Cd have a significant impact upon lattice parameter, ground state properties etc [40–42]. Due to the transformed phase nearest neighbour distance, bond length would get change and thereafter modify the *d* electrons sub-band states. Oxygen vacancies have a deep impact upon the complete spectral profile and that has been well described by *I.N. Demchenko et al* [43] theoretically. In present scenario for CZ900 thin film has been annealed at 900°C for one hour, Therefore, it is expected that density of oxygen vacancies will be lesser for CZ900 thin film which will affect the O *K* edge spectral profile and implement observable modification compared to CZ700 thin film.

The B and C region beyond the absorption edge are generated due to interference effect of multiple scattering signal coming from second, third or higher coordination. However, significant change has not been observed in the C region. This observation is quite matching fine with theoretical calculation by Demchenko *et al* [43]. Calculation predicts if one keeps on increasing cluster size it will not generate any new features in the spectra at higher coordination.

**Conclusion:**

In conclusion, high temperature driven B1 to B2 structural PT in 40% Cd doped ZnO has been reported. In fact X-ray diffraction spectra and complementary Raman spectroscopic results clearly revealed the evolution of B2 phase annealed from ~700℃ to ~900℃. Structural refinement for Fullprof Suit Toolbar reflects the reduction in partial volume fraction for the transformed phase, suggesting the presence of chemical pressure inside the lattice which in turn eventually responsible for PT. In addition, SEM micrographs reflect grain growth, followed by agglomeration of nanocrystallites along with void formation with increasing annealing temperature. Shifting in the absorption edge position in oxygen *K*-edge spectra is found to observed, signifying the modification in hybridization and sub band states with PT. Moreover, for clear understanding and mechanism responsible for development of such chemical pressure inside the lattice, detailed study of nearest neighbour distance, coordination number is required which is now under investigation and reports elsewhere. Phonon density of states calculation, extended region XAS for probing Cd *K* edge will be quite crucial for investigating the transformation kinetics. Overall for the first time such kind of chemical pressure induced PT has been reported for solgel derived $Cd_{0.4}Zn_{0.6}O$ pseudo binary alloy system experimentally in this work.

**Acknowledgement:**

Authors are gratified to director of IUAC, director of UGC-CSR Indore, director of RRCAT for making available the experimental facilities which have cherished the research work presented in this manuscript. One of the author (Dr. A. Das) is obliged to Dr. Fouran Singh and Dr. S.K. Gautam for discussions, Dr. D. K. Shukla and Mr. G. R. Umapathy for SXAS and RBS measurements respectively. One of the authors (Dr. A. Das) is also grateful to university grant commission (UGC) for providing fellowship (Grant number F.2-91/1998(SA-1)). Deaprtment of Science and technology (DST), Govt. of India are acknowledged for granting Science and Engineering Research Board (SERB) project (SB/EMEQ-122/2013).

**Table caption:**

Calculated values of lattice parameters, volume phase fraction from structural refinement with FULL PROF SUITE software for pure ZnO, CZ700 and CZ900 thin films

**Figure captions:**

1. Rutherford Backscattering Spectra and Rump simulation for CZ700 and CZ900 thin films
2. X-ray diffraction pattern for CZ700, CZ750, CZ800, CZ850, CZ850, CZ900 thin films (a) and refinement for PZ (b), CZ700 (c) and CZ900 (d) thin films
3. Raman Spectra for CZ700, CZ800, CZ900 thin films
4. SEM images for CZ700, CZ800, CZ900 thin films in 100 nm and 1 μm scale
5. Oxygen *K* edge spectra for CZ700 and CZ900 thin films
6. Peak fitting with Athena XAS software package for CZ700 and CZ900 thin film

| **Sample Name** | **Space Group** | **Lattice Parameters** | **Volume** | **GOF** |
|---|---|---|---|---|
| **PZnO** | P63mc (Hexagonal) | a=b=3.24643(5) Å $\alpha = \beta = 90^o$ c=5.20076(8) Å $\gamma = 120^o$ | Vol=47.469 Å$^3$ | $\chi^2$=1.23 |
| **CZ700** | Fm3m (cubic) | a=b=c=4.68209 (3) A$^o$ $\alpha = \beta = \gamma = 90^o$ | Vol=102.641Å$^3$ | $\chi^2$=1.88 |
| | P63mc (Hexagonal) | a= b=3.24953 (2) Å $\alpha = \beta = 90^o$ c= 5.20415 (4) Å $\gamma = 120^o$ | Vol=47.591 Å$^3$ | |
| **CZ900** | Pm-3m (cubic) | a=b=c= 2.78241 (6) Å $\alpha = \beta = \gamma = 90^o$ | Vol=21.541 Å$^3$ | $\chi^2$=3.54 |
| | P63mc (Hexagonal) | a= b= 3.26361 (3) Å $\alpha = \beta$=90$^o$ c= 5.22686 (7) Å $\gamma = 120^o$ | Vol=48.214 Å$^3$ | |

**Table -1**

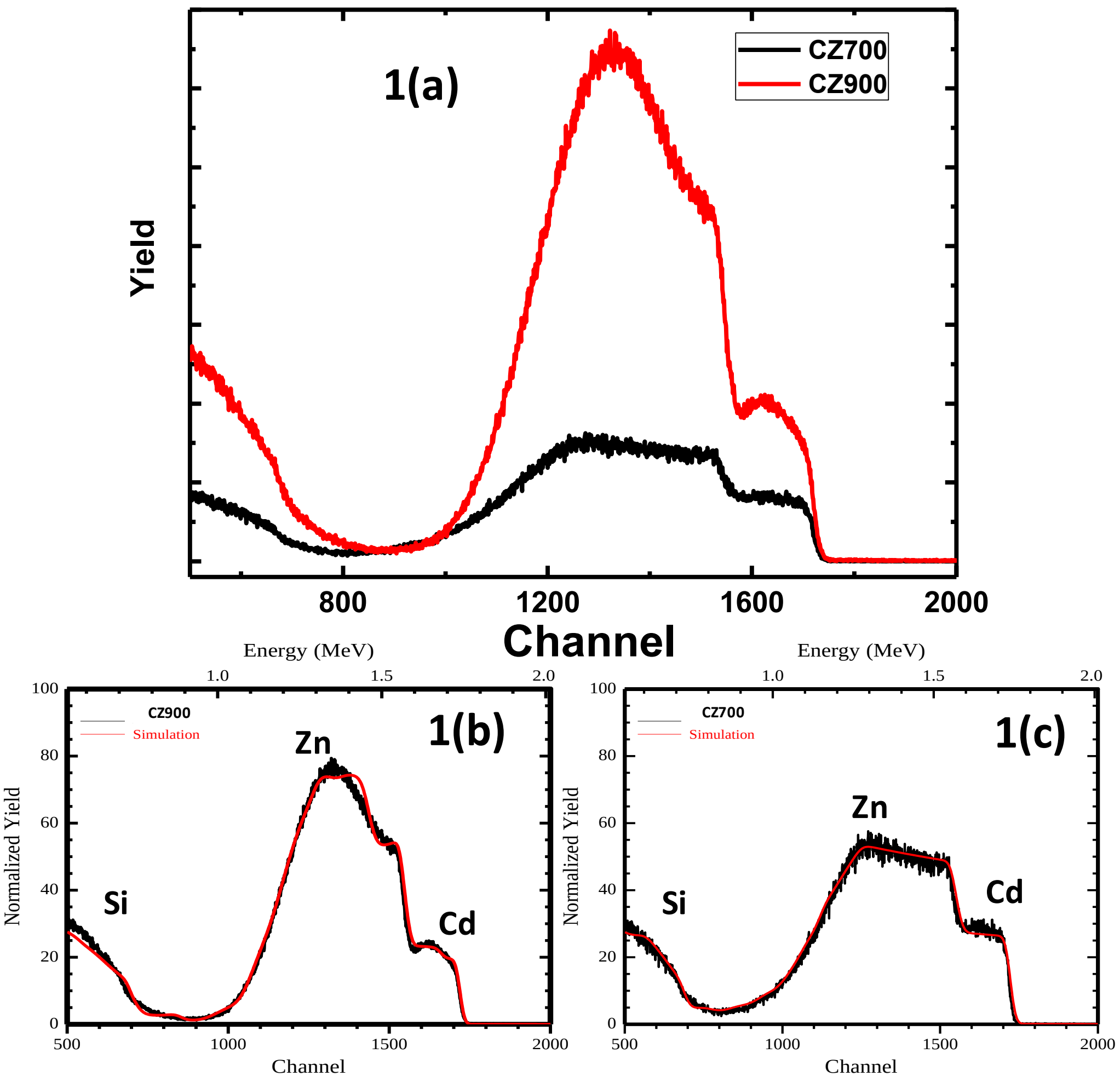

**Figure 1**

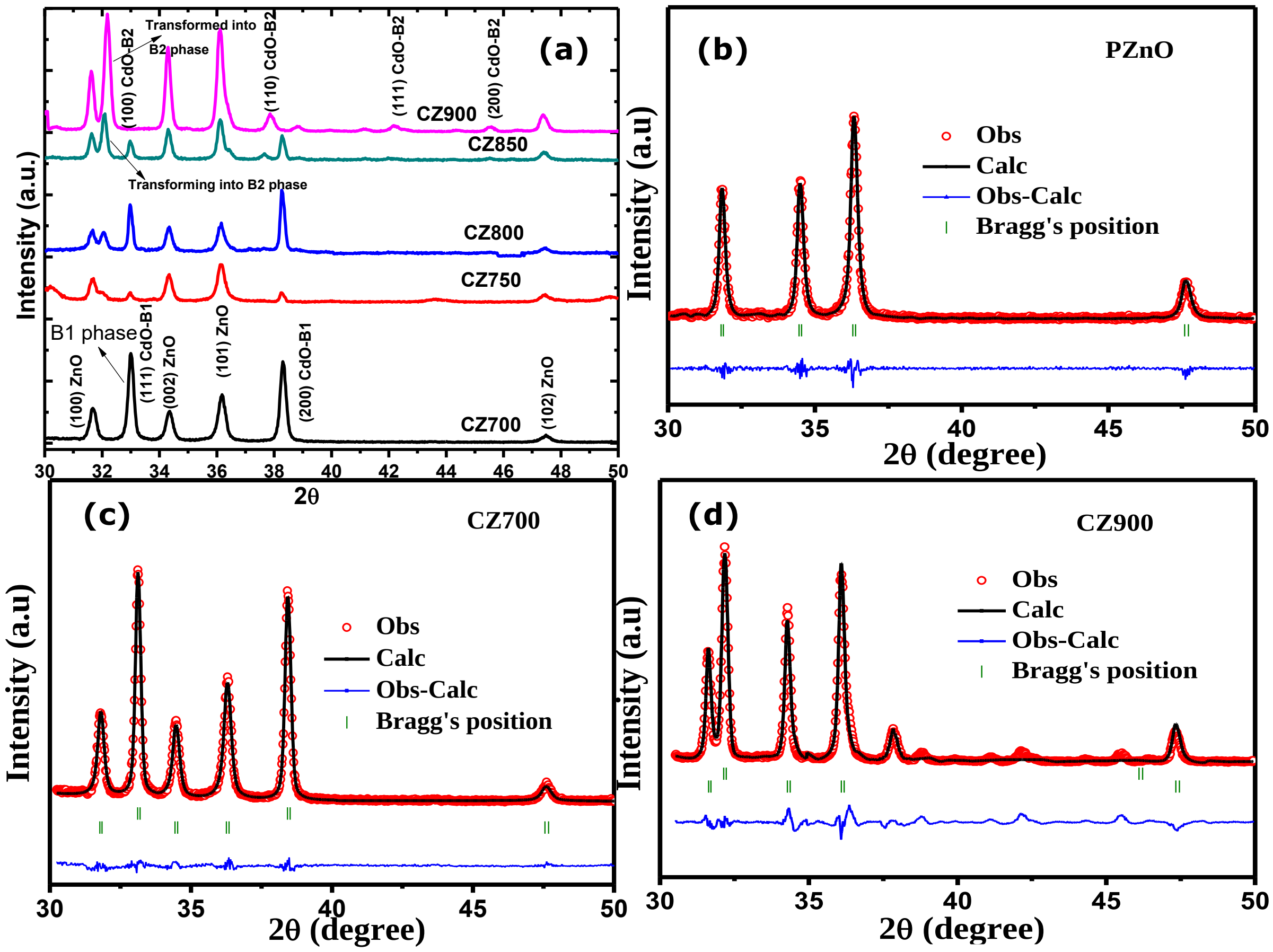

**Figure 2**

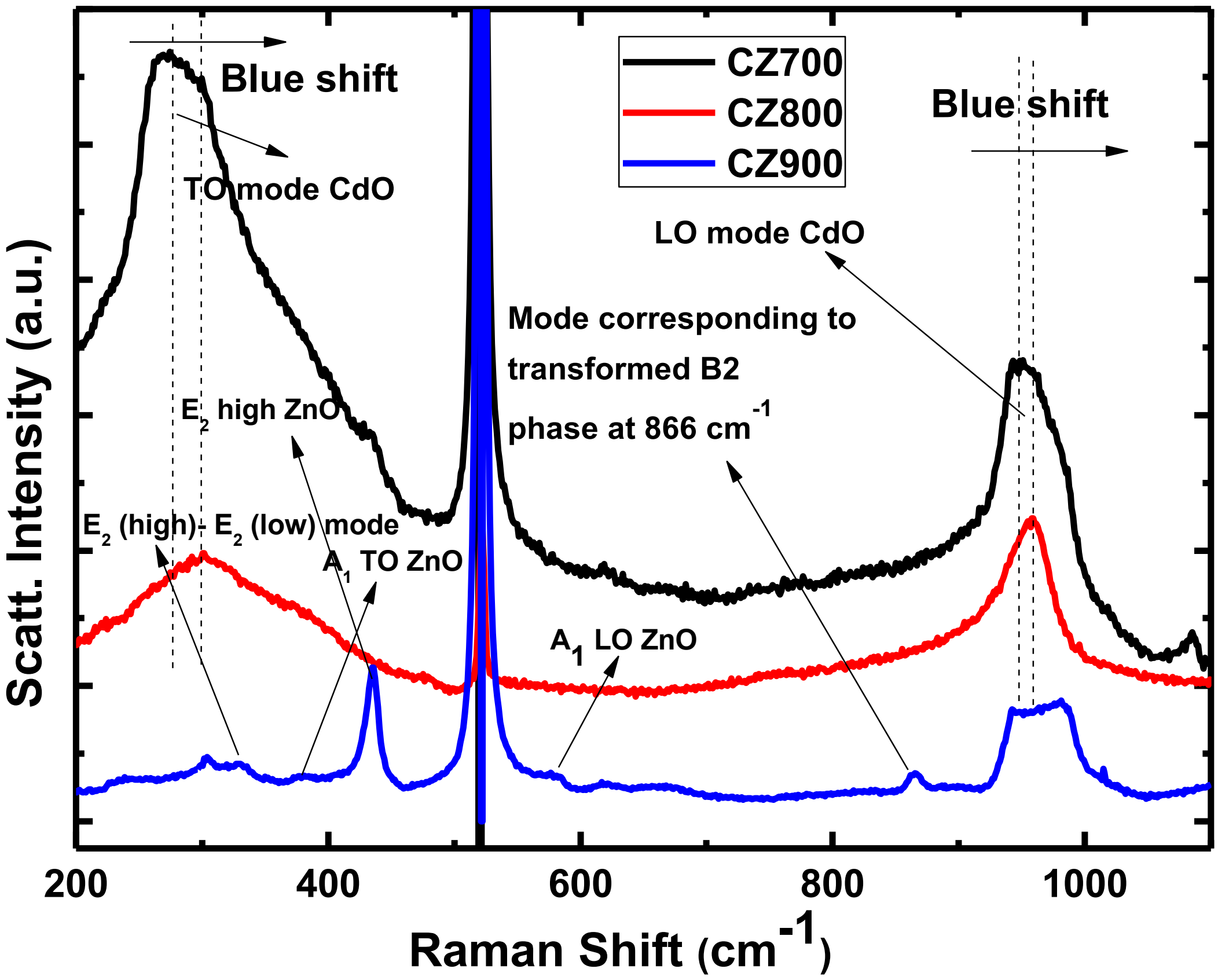

**Figure 3**

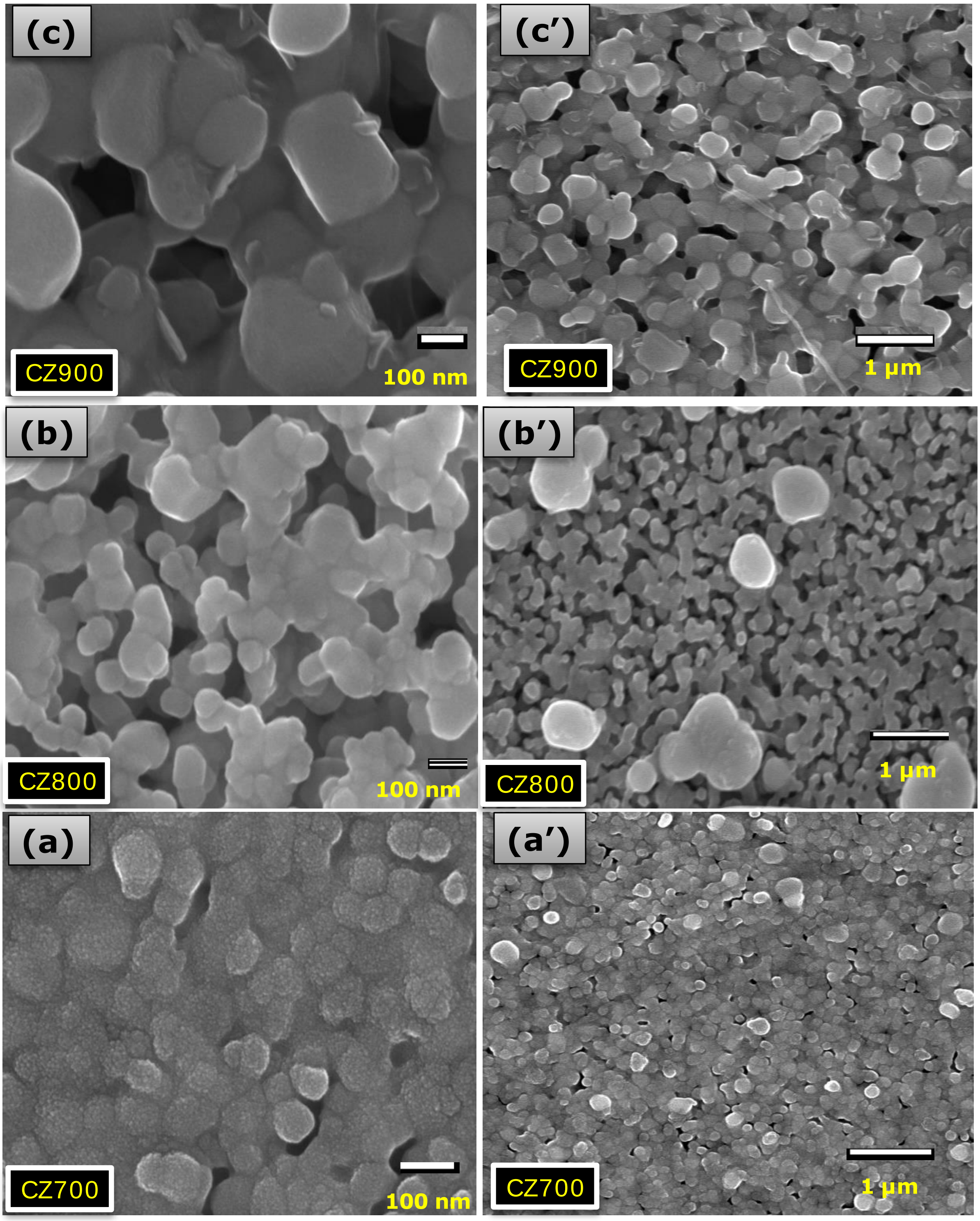

**Figure 4**

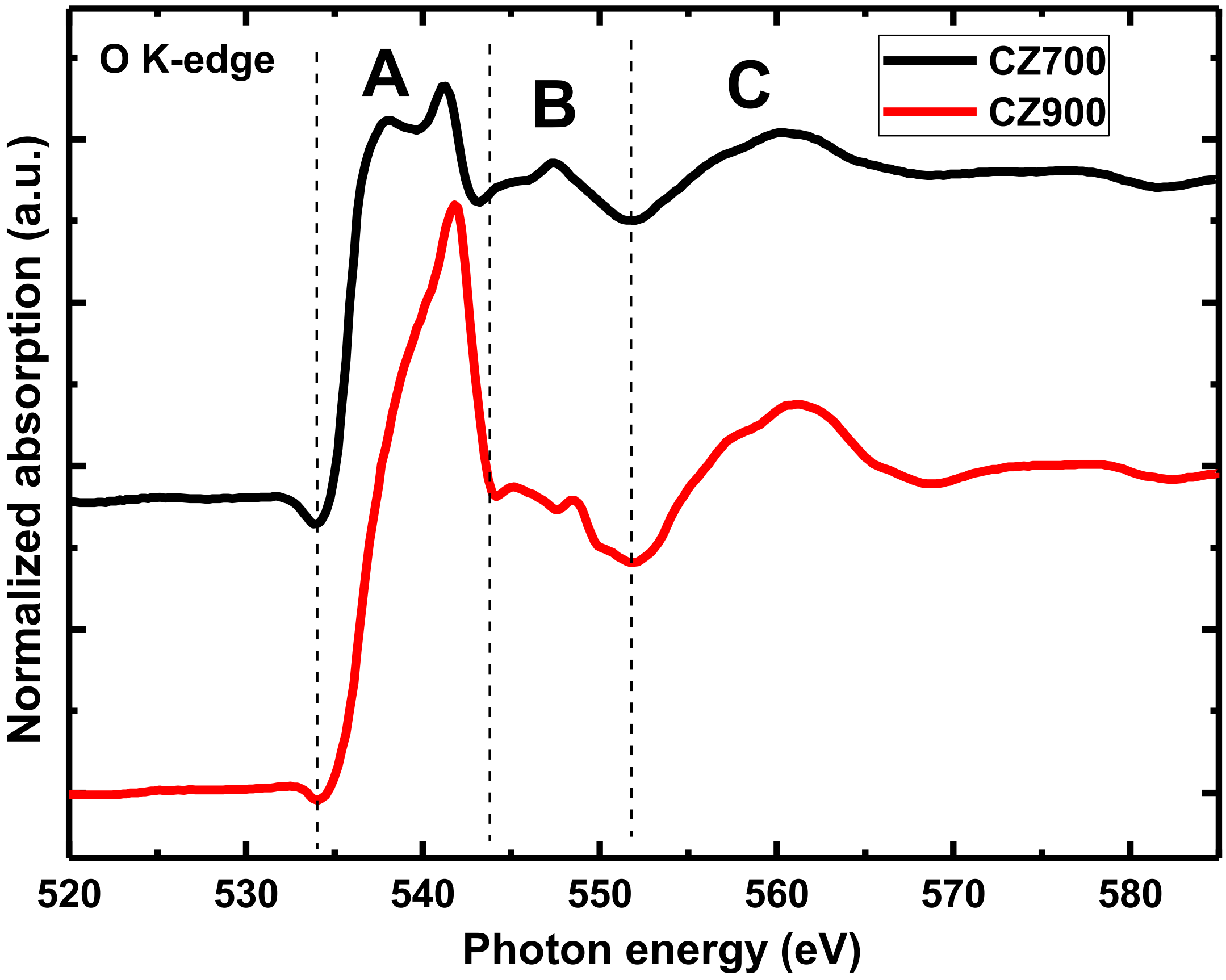

**Figure 5**

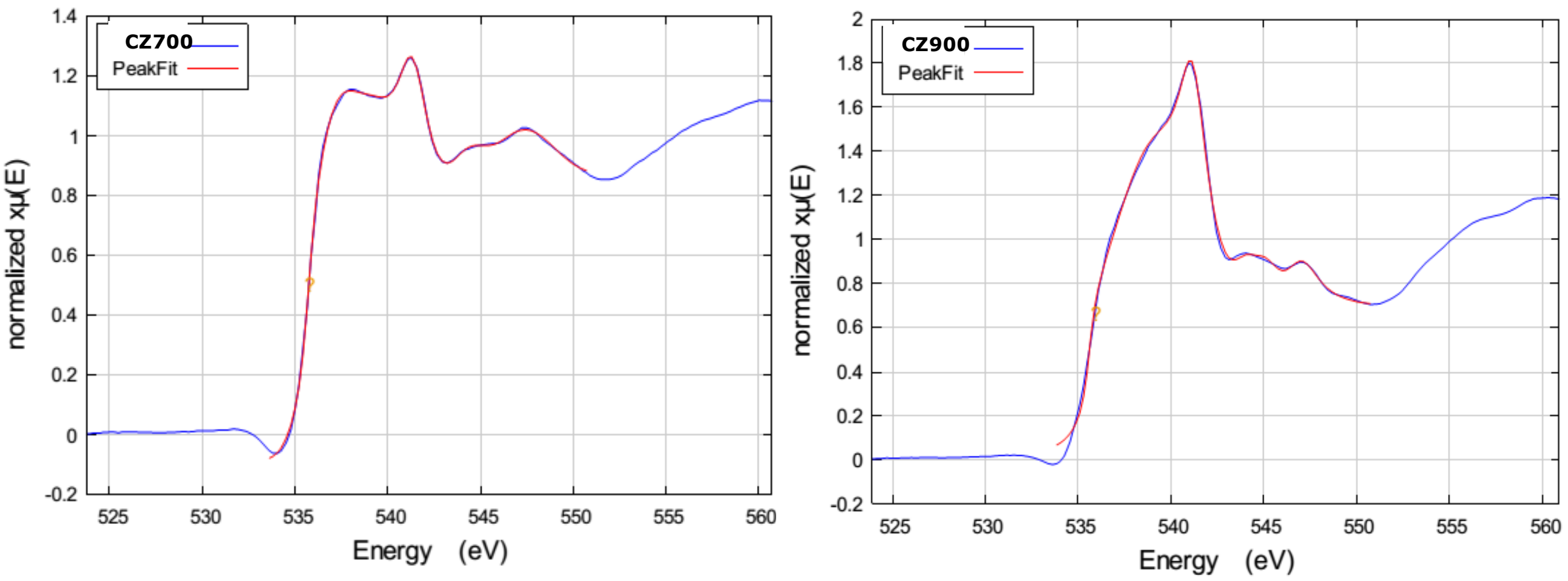

**Figure 6**